\documentclass[prd,aps,showpacs,nofootinbib,tightenlines]{revtex4}
\usepackage{mathrsfs}
\usepackage{amsmath}
\usepackage{amssymb}
\usepackage{epsfig}
\usepackage{graphicx}
\usepackage{booktabs}
\usepackage{multirow}
\usepackage{subfigure}
\usepackage{bm}
\usepackage{times}
\usepackage{braket}
\usepackage{color}
\usepackage{slashed}
\usepackage{hyperref}
\definecolor{Red}{rgb}{1.,0.,0.}

\definecolor{Blue}{rgb}{0.,0.,1.}

\definecolor{nicered}{rgb}{0.7,0.1,0.1}
\definecolor{nicegreen}{rgb}{0.1,0.5,0.1}
\hypersetup{colorlinks,citecolor=nicegreen,linkcolor=nicered}
\def \cpc{ Chin. Phys. C  }

\def \epjc{ Eur. Phys. J. C }
\def \ijmpa{ Int. J. Mod. Phys. A }

\def \npb{  Nucl. Phys. B }
\def \plb{  Phys. Lett. B }
\def \ppnp{ Prog.Part. $\&$ Nucl. Phys. }
\def \prd{  Phys. Rev. D }
\def \prl{  Phys. Rev. Lett.  }

\def \jhep{ JHEP }
\begin{document}

\newcommand{\beq}{\begin{eqnarray}}
\newcommand{\eeq}{\end{eqnarray}}
\newcommand{\non}{\nonumber\\ }

\title{Quasi-two-body decays $B_{(s)}\to K^*(892)h\to  K\pi h$ in perturbative QCD approach}

\author{Ya Li$^{1}$}               \email{liyakelly@163.com}
\author{Wen-Fei Wang$^2$}       \email{wfwang@sxu.edu.cn}
\author{Ai-Jun Ma$^3$}         \email{theoma@163.com}
\author{Zhen-Jun Xiao$^{4}$}  \email{xiaozhenjun@njnu.edu.cn}
\affiliation{$^1$ Department of Physics, College of Sciences, Nanjing Agricultural University, Nanjing, Jiangsu 210095, P.R. China}
\affiliation{$^2$ Institute of Theoretical Physics, Shanxi University, Taiyuan, Shanxi 030006, P.R. China}
\affiliation{$^3$ Department of Mathematics and Physics, Nanjing Institute of Technology, Nanjing, Jiangsu 211167, P.R. China}
\affiliation{$^4$ Department of Physics and Institute of Theoretical Physics,
                          Nanjing Normal University, Nanjing, Jiangsu 210023, P.R. China}
\date{\today}

\begin{abstract}
We extend our recent works on the $P$-wave two-pion resonant contributions to the kaon-pion cases in the hadronic
charmless $B$ meson decays by employing the perturbative QCD approach. The concerned decay modes are analysed
in the quasi-two-body framework by parameterizing the kaon-pion distribution amplitude $\Phi_{K \pi}^{\rm P}$, which
contains the final state interactions between the kaon and pion in the resonant region. The relativistic Breit-Wigner formula
for the $P$-wave resonant state $K^*(892)$ is adopted to parameterize the time-like form factor $F_{K\pi}$.
We calculate the $CP$-averaged branching ratios and direct $CP$-violating asymmetries of the quasi-two-body decays
$B_{(s)}\to  K^*(892) h\to  K\pi h$, with $h=(\pi, K)$, in this work. It is shown that the agreement of theoretical results with the
experimental data can be achieved, through which Gegenbauer moments of the $P$-wave kaon-pion distribution amplitudes
are determined.
The predictions in this work will be tested by the precise data from the LHCb and the future Belle II experiments.
\end{abstract}

\pacs{13.25.Hw, 12.38.Bx}

\maketitle
\section{Introduction}
Three-body hadronic $B$ meson decays are a rich field for the experiments and theoretical studies.
These decay processes offer one of the best tools for the analyses of direct $CP$ violation and also provide a testing ground for
the dynamical models of the strong interaction. Strong dynamics in a three-body $B$ meson decay is much more complicated
than that in a two-body case, the three-body processes receive nonresonant and resonant contributions, as well as the significant
final-state interactions~\cite{prd89-094013,1512-09284,89-053015}.
The nonresonant contributions have been studied with the
method of heavy meson chiral perturbation theory (HMChPT)~\cite{prd46-1148,prd45-2188,prd55-5851,plb280-287} valid in the soft meson limit in Ref.~\cite{prd94-094015}.
The exponential factor $e^{-\alpha_{\rm NR} p_B\cdot (p_i+p_j)}$ is introduced so that the HMChPT results are recovered in the soft meson
limit where $p_i, p_j \to 0$.
In addition to the nonresonant background, it is urgent to study the resonant contributions which are, in most cases, the dominant
part of a three-body decay process. Analyses of the three-body decays utilizing the Dalitz plots~\cite{dalitz-plot1,dalitz-plot2} enable us
to investigate the properties of various scalar, vector and tensor resonant states with the isobar model~\cite{123-333,prd11-3165}
in terms of the usual Breit-Wigner model~\cite{BW-model}.

On the theoretical side, no proof of factorization has been done for the decays of the $B$ meson into three final state mesons.
As a first step, however, we can restrict ourselves to the specific kinematical configurations, in which two energetic final state
mesons almost collimating to each other. For such topologies, the three-body interactions are expected to be suppressed
strongly due to power counting rules~\cite{prd79-094005}.
In such quasi-two-body region of phase space, the obvious generalization of the factorization theorem for two-body decays applies, and in the region where all invariant masses are large, factorization has been explicitly shown at the leading non-trivial order in Refs.~\cite{npb675-333,beneke}.
It's reasonable for us to assume the validity of the factorization for these quasi-two-body $B$ decays.
In the ``quasi-two-body" mechanism, the two-body scattering and all possible interactions between the two involved particles are included
but the interactions between the bachelor particle and the daughter mesons from the resonance are neglected.
Substantial progress on three-body hadronic $B$ meson decays by means of symmetry principles has been
made for example in Refs.~\cite{prd72-094031,plb727-136,prd72-075013,prd84-056002,plb726-337,plb728-579,IJMPA29-1450011,prd91-014029}.
The QCD-improved factorization~\cite{npb675-333} has also been widely applied to the studies of the three-body hadronic $B$ meson
decays in Refs.~\cite{npb899-247,plb622-207,prd74-114009,prd79-094005,APPB42-2013,prd76-094006,prd88-114014,
prd94-094015,prd89-094007,prd87-076007,prd81-094033}.
The detailed factorization properties of the $B^+\to\pi^+\pi^+\pi^-$
in different regions of phase space were investigated in Ref.~\cite{npb899-247}.
The $CP$ violations and the contributions of the strong kaon-pion interactions have been investigated in the $B \to K\pi\pi$
decays utilizing an approximate construction of relevant scalar and vector form factors in Ref.~\cite{prd79-094005}.
In Ref.~\cite{APPB42-2013}, the authors studied the decays of $B^\pm \to \pi^\pm \pi^\mp \pi^\pm$ within a quasi-two-body
QCD factorization approach and introducing the scalar and vector form factors for the $S$ and $P$ waves, as well as a relativistic
Breit-Wigner (RBW) formula for the $D$ wave to describe the meson-meson final state interactions.

The perturbative QCD (PQCD) factorization approach has been employed in Refs.~\cite{plb561-258,prd70-054006,prd97-034033,prd89-074031,1803-02656},
where the strong dynamics between the two final state hadrons in resonant regions are factorized into a new non-perturbative input, the two-hadron distribution amplitudes (DAs)
$\Phi_{h_1h_2}$~\cite{MP,MT01,MT02,MT03,MN,Grozin01,Grozin02}.
Both nonresonant and resonant contributions can be accommodated into this new input in PQCD factorization approach.
In the PQCD approach, we have studied the $S$-wave resonance contributions to the decays of $B^0_{(s)}$ mesons into
a charmonium meson plus pion-pion (koan-pion) pair~\cite{prd91-094024,epjc76-675,cpc41-083105,epjc77-199,prd97-033006},
the $P$-wave resonance contributions to the decays $B \to P (\rho,\rho(1450),\rho(1700)) \to P \pi\pi$~\cite{plb763-29,prd95-056008,prd96-036014}, $B_{(c)} \to D (\rho,\rho(1450),\rho(1700)) \to D \pi\pi$~\cite{npb923-54,1708-01889,1710-00327} and $B \to \eta_c(1S,2S) (\rho,\rho(1450),\rho(1700)) \to \eta_c(1S,2S)\pi\pi$~\cite{npb924-745}, as well as the $D$-wave resonant contributions to the decays $B \to P f_2(1270) \to P \pi\pi$~\cite{ly2018}.
All these works indicate that the PQCD factorization approach is universal for exclusive hadronic three-body $B$ meson decays.

The measurements for the branching ratios and $CP$ violating asymmetries for $B \to K \pi \pi $
and other decay modes have been reported by {\it BABAR} ~\cite{prd70-092001,prd72-072003,prd80-112001,prd79-072006,prd78-052005,prd83-112010},
Belle~\cite{prd75-012006,prl96-251803,prd71-092003,prd79-072004}
and LHCb Collaborations~\cite{epjc73-2373,prl111-101801,jhep10-143,prd90-112004,prl112-011801,prd95-012006,jhep07-021}.
These three-body decays are known experimentally to be dominated by the low energy resonances on $\pi\pi$, $KK$ and $K\pi$
channels on the Dalitz plots. In this work, we shall extend our recent works on the $P$-wave two-pion resonant contributions to
the kaon-pion cases. Motivated by the recent detailed Dalitz plot analyses of $K\pi$ invariant mass spectrum by
{\it BABAR}~\cite{prd80-112001,prd83-112010,prd82-011502,prl97-201802,prd78-012004,prd84-092007},
Belle~\cite{prd75-012006,prd79-072004,prd75-092002,prd75-092005,prl96-251803,prd90-072009,plb599-148},
CLEO~\cite{prl85-520,prl85-2881,prl89-251801} and LHCb~\cite{NJP16-123001,jhep01-012} Collaborations, we will calculate the decay
modes $B \to K\pi h$, where $h$ is the light pseudoscalar pion or kaon, and study the $K\pi$ pair originating from a vector
quark-antiquark state, while other partial waves are beyond the scope of the present work.

The relevant Feynman diagrams are the same as Fig.~1 in the Ref.~\cite{prd95-056008}.
The $P$-wave contributions are parameterized into the time-like vector form factors involved in the kaon-pion DAs.
We adopt the RBW line shape for the $P$-wave resonance $K^*(892)$ to parameterize the time-like form
factors~\cite{prd83-112010}. Throughout the remainder of the paper, the symbol $K^*$ is used to denote the $K^*(892)$ resonance.
By employing the kaon-pion DAs, the $P$-wave contributions to the related three-body $B$ meson decays can be
simplified into quasi-two-body processes $B \to K^* h\to K\pi h$.

As is well known, the QCD-improved factorization (QCDF)~\cite{npb675-333,prl83-1914,npb591-313,prl96-141801,npb832-109}, the perturbative QCD (PQCD) factorization
approach~\cite{prd63-074009,plb504-6,ppnp51-85} and the soft-collinear-effective theory (SCET)~\cite{prd70-054015,prd74-034010,npb692-232,prd72-098501,prd72-098502} are the three popular factorization approaches to deal with the hadronic $B$ meson decays.
For most $B \to h_1 h_2$ decay channels, the theoretical predictions obtained by using these different
factorization  approaches agree well with each other and also are well consistent with the data within errors.
QCDF and SCET are based on the collinear factorization, in which $B$ meson transition form factors contain the end-point singularity.
That is why soft form factors need to be introduced in these approaches.
PQCD is based on the $k_T$ factorization, in which the implementation of the Sudakov resummation suppresses the small $k_T$ region.
It has been shown that $k_T^2$ is of order $m_b\Lambda_{QCD}$ in Ref.~\cite{prd65-014007}.
The end-point singularity is then smeared, and the form factors are factorizable.
That is, the different power countings for the parton $k_T$ lead to different factorization formalisms.
As to the other modes, the hard-collinear modes correspond to the hard kernels in PQCD, and the collinear modes correspond to the DAs in PQCD.
The calculation of the infrared logarithms in QCD and in $k_T$-dependent DAs for the $B \to \pi$ form factors and their cancellation have been done at one-loop level in Ref.~\cite{prd85-074004}.
One can see that the $k_T$-dependent next-to-leading-order hard kernel for the $B \to \pi$ transition form factors is infrared-finite.
Therefore, the form factors in $B \to \pi$ transition are factorizable.
The all-order proof for the $k_T$ factorization of the $B \to \pi$ form factors has been done in Ref.~\cite{prd67-034001}.

With the introduction of a two-meson DAs, the LO diagrams for three-body hadronic $B$ meson decays are reduced to those for quasi-two-body decays.
The hard kernel $H$ describes the dynamics of the strong and electroweak interactions in three-body hadronic decays in a similar way as the one for the two-body $B\to h_1 h_2$ decays.
The $\Phi_B$ and $\Phi_{h_3}$ are used to describe the wave functions for the $B$ meson and the bachelor particle $h_3$, which absorb
the non-perturbative dynamics in the process.
The $\Phi_{h_1h_2}$ is the two-hadron ($K$ plus $\pi$ in this work) distribution amplitude, which describes the structure of the final-state $K$-$\pi$ pair.
As a result, one can describe the typical PQCD factorization
formula for a $B\to h_1h_2h_3$  decay amplitude as the form of~\cite{plb561-258,prd70-054006},
 \begin{eqnarray}
\mathcal{A}=\Phi_B\otimes H\otimes \Phi^{\text{$P$-wave}}_{h_1h_2}\otimes\Phi_{h_3}.
\end{eqnarray}

This paper is organized as follows.
In Sec.~II, we give a brief introduction for the theoretical framework.
The numerical results, some discussions and the conclusions will be given in last two sections.
The factorization formulas for the relevant three-body decay amplitudes are collected in the Appendix.

\section{FRAMEWORK}\label{sec:2}

In the light-cone coordinates, we let the kaon-pion pair and the final-state $h$ move along the direction of
$n=(1,0,0_{\rm T})$ and $v=(0,1,0_{\rm T})$, respectively, in the rest frame of the $B$ meson.
The kinematic variables of the decay $B(p_B) \to (K\pi)(p)h(p_3)$ can be chosen as
\begin{eqnarray}\label{mom-pBpp3}
p_{B}=\frac{m_{B}}{\sqrt2}(1,1,0_{\rm T}),~\quad p=\frac{m_{B}}{\sqrt2}(1,\eta,0_{\rm T}),~\quad
p_3=\frac{m_{B}}{\sqrt2}(0,1-\eta,0_{\rm T}),
\end{eqnarray}
where $m_{B}$ is the mass of $B$ meson, the variable $\eta$ is defined as $\eta=\omega^2/m^2_{B}$,
the invariant mass squared $\omega^2=p^2$ for the kaon-pion pair.
If we choose $\zeta=p^+_1/p^+$ as kaon momentum fraction, the kaon momentum $p_1$ and pion momentum $p_2$ can be written as
\begin{eqnarray}
 p_1=(\zeta \frac{m_B}{\sqrt{2}}, (1-\zeta)\eta \frac{m_B}{\sqrt{2}}, p_{1\rm T}),~\quad
 p_2=((1-\zeta)\frac{m_B}{\sqrt{2}}, \zeta\eta \frac{m_B}{\sqrt{2}}, p_{2\rm T}).
\end{eqnarray}
We employ $x_B, z, x_3$ to denote the momentum fraction of the positive quark in each meson,
$k_{BT}, k_{\rm T}, k_{3{\rm T}}$ is assigned to the transverse momentum of the positive quark, respectively.
The momentum $k_B$ of the spectator quark in the $B$ meson, the momentum $k$ for the resonant state $K^*(892)$ and $k_3$ for the final-state $h$ are of the form of
\begin{eqnarray}
k_{B}&=&\left(0,x_B \frac{m_{B}}{\sqrt2} ,k_{BT}\right),\quad k= \left( \frac{m_{B}}{\sqrt2}z,0,k_{\rm T}\right),\quad
k_3=\left(0,(1-\eta)x_3 \frac{m_B}{\sqrt{2}},k_{3{\rm T}}\right), \label{mom-B-k}
\end{eqnarray}
The momentum fractions $x_{B}$, $z$ and $x_3$ run from zero to unity.

The $P$-wave kaon-pion DAs are introduced in analogy with the case of two-pion ones~\cite{plb763-29} if ignoring the masses of the kaon and pion mesons,
\begin{eqnarray}
\Phi_{K\pi}^{\text{$P$-wave}}=\frac{1}{\sqrt{2N_c}}[{ p \hspace{-2.0truemm}/ } \phi_0(z,\zeta,\omega^2)+\omega \phi_{s}(z,\zeta,\omega^2)
+\frac{{p\hspace{-1.5truemm}/}_1{p\hspace{-1.5truemm}/}_2
  -{p\hspace{-1.5truemm}/}_2{p\hspace{-1.5truemm}/}_1}{\omega(2\zeta-1)} \phi_t(z,\zeta,\omega^2)] \;.
\label{eq:phifunc}
\end{eqnarray}

The expansions of the nonlocal matrix elements for the vector, scalar and tensor spin projectors up to twist-3 are listed below similar to our recent work~\cite{1809-04754},
\begin{eqnarray}
\langle K(p_1)\pi(p_2)|\bar{q}_1(y^-)\gamma_{\mu}q_2(0)|0\rangle&=&(p_1-p_2)_{\mu}\int_0^1
dz e^{izP\cdot y}\phi_0(z,\omega), \label{eq:ff1}\\
\langle K(p_1)\pi(p_2)|\bar{q}_1(y^-) I q_2(0)|0\rangle&=&\omega \int_0^1
dze^{izP\cdot y}\phi_s(z,\omega), \label{eq:ff2}\\
\langle K(p_1)\pi(p_2)|\bar{q}_1(y^-)\sigma_{\mu\nu}q_2(0)|0\rangle&=&-i\frac{(p_{1\mu}p_{2\nu}-p_{1\nu}p_{2\mu})}{\omega}\int_0^1
dze^{izP\cdot y}\phi_t(z,\omega),\label{eq:ff3}
\end{eqnarray}
with the quark content $q_1=s, q_2=u(d)$ or $q_1=u(d), q_2=s$.
The DA $\phi_0$ is the twist-2 component, while the DAs $\phi_s,\phi_t$ are the twist-3 ones.
Following the steps of $S$-wave kaon-pion resonance~\cite{prd97-033006,plb730-336}, it is worthwhile to stress that the $P$-wave kaon-pion system has similar DAs as the ones for a light vector meson, but we replace the vector decay constants with the time-like form factor:
\begin{eqnarray}
\phi_0&=&\frac{3F_{K\pi}(s)}{\sqrt{2N_c}} z
(1-z)\left[1+a_{1K^*}^{||}3(2z-1)+
a_{2K^*}^{||}\frac{3}{2}(5(2z-1)^2-1)\right]P_1(2\zeta-1)\;,\label{eq:phi1}\\
\phi_s&=&\frac{3F_s(s)}{2\sqrt{2N_c}}(1-2z)P_1(2\zeta-1) \;, \label{eq:phi2}\\
\phi_t&=&\frac{3F_t(s)}{2\sqrt{2N_c}}(2z-1)^2P_1(2\zeta-1) \;,
\label{eq:phi3}
\end{eqnarray}
where the Legendre polynomial $P_1(2\zeta-1)=2\zeta-1$ and the Gegenbauer moments $a_{1K^*}^{||}$ and $a_{2K^*}^{||}$ will be regarded as free parameters and determined in this work.
The time-like form factors $F_{K\pi}(s), F_s(s), F_t(s)$ define the normalization of the $K\pi$ two-meson distribution amplitudes.
Note that the hadronic matrix element in Eq.~(\ref{eq:ff2}) vanishes
for the local operator with $y^-=0$, namely, as the DA $\phi_s$ is
integrated over the parton momentum fraction $z$.
It implies that the scalar form factor $F_s$ can be defined only via a nonlocal matrix
element.
In principle, $F_s$ and $F_t$ should be different.
However, we are not able to distinguish them currently because of limited data.
Thus, we supposed that they are equal.
Following Ref.~\cite{plb763-29}, we also assume that
\begin{eqnarray}
F_s(s) = F_t(s) \approx (f_{K^*}^T/f_{K^*}) F_{K\pi}(s).
\end{eqnarray}
with $f_{K^*}=0.217 \pm 0.005 {\rm GeV}, f^T_{K^*}=0.185 \pm 0.010 {\rm GeV}$~\cite{prd76-074018}.

For the narrow resonance $K^*$, we adopt the RBW line shape for the $P$-wave resonance $K^*$ to parameterize the time-like
form factors $F_{K\pi}(s)$, which is widely adopted in the experimental data analyses.
The explicit expressions are in the following form~\cite{prd83-112010},
\begin{eqnarray}
F_{K\pi}(s)&=&\frac{m_{K^*}^2}{m^2_{K^*} -s-im_{K^*}\Gamma(s)},
\end{eqnarray}
with the kaon-pion invariant mass squared $s=\omega^2=m^2(K\pi)$.

Here, the mass-dependent width $\Gamma(s)$ is defined by
\begin{eqnarray}
\Gamma(s)&=&\Gamma_{K^*}\frac{m_{K^*}}{\sqrt{s}}\left(\frac{|\overrightarrow{p_1}|}{|\overrightarrow{p_0}|}\right)^3,
\end{eqnarray}
where $m_{K^*}$ and $\Gamma_{K^*}$ are the pole mass and width of resonance state $K^*$ respectively.
The $|\overrightarrow{p_1}|$ is the momentum vector of the resonance decay product measured in the resonance rest frame, while $|\overrightarrow{p_0}|$ is the value of $|\overrightarrow{p_1}|$ when $\sqrt{s}=m_{K^*}$.
The explicit expression of kinematic variables $|\overrightarrow{p_1}|$ is
\begin{eqnarray}
|\overrightarrow{p_1}|=\frac{\sqrt{\lambda(\omega^2,m_K^2,m_{\pi}^2)}}{2\omega},
\end{eqnarray}
with the kaon (pion) mass $m_K$ ($m_{\pi})$ and the K$\ddot{a}$ll$\acute{e}$n function $\lambda (a,b,c)= a^2+b^2+c^2-2(ab+ac+bc)$.


\section{Numerical results}\label{sec:3}

\begin{figure}[tbp]
\centerline{\epsfxsize=7.5cm \epsffile{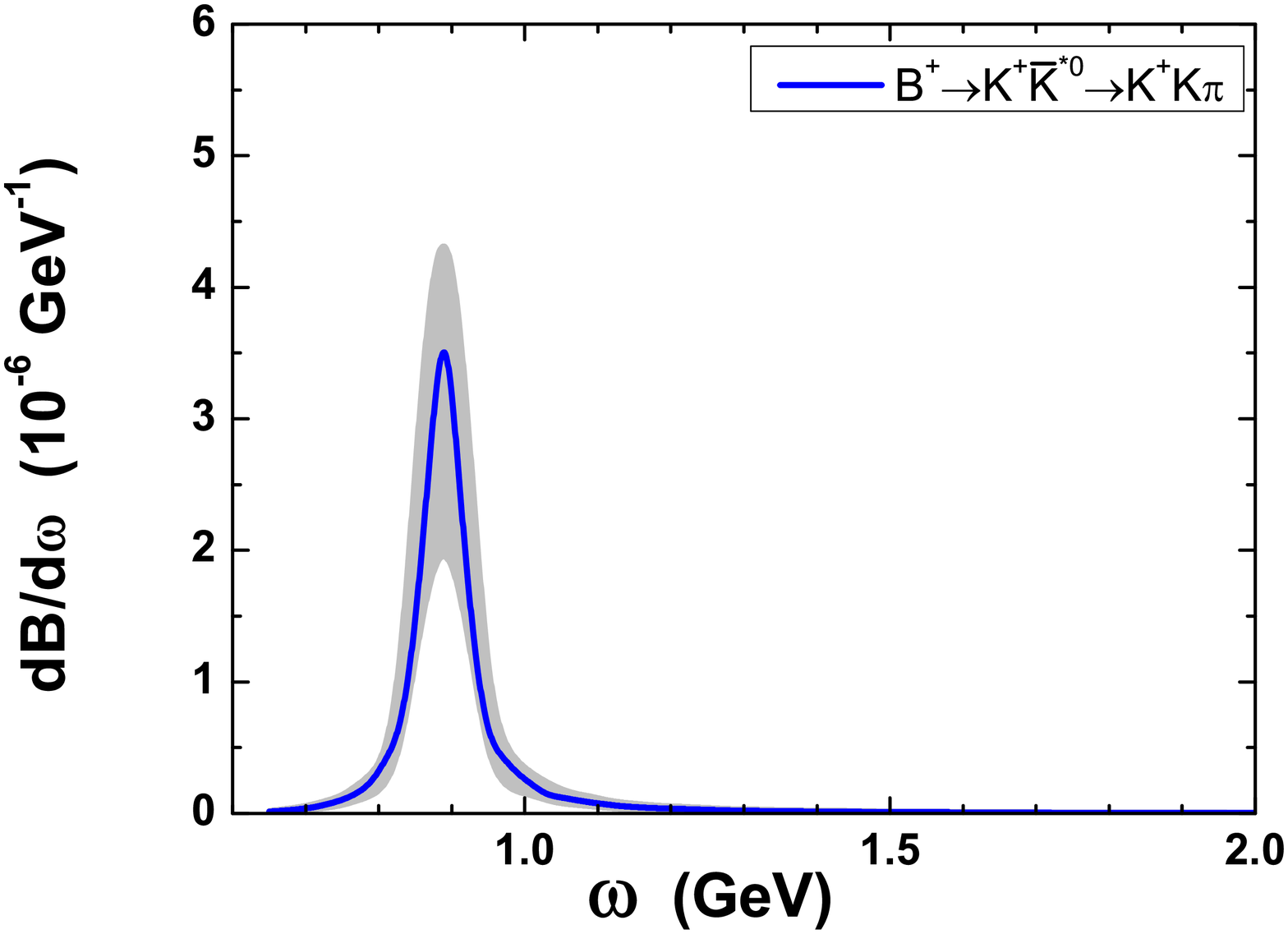}
            \epsfxsize=7.5cm \epsffile{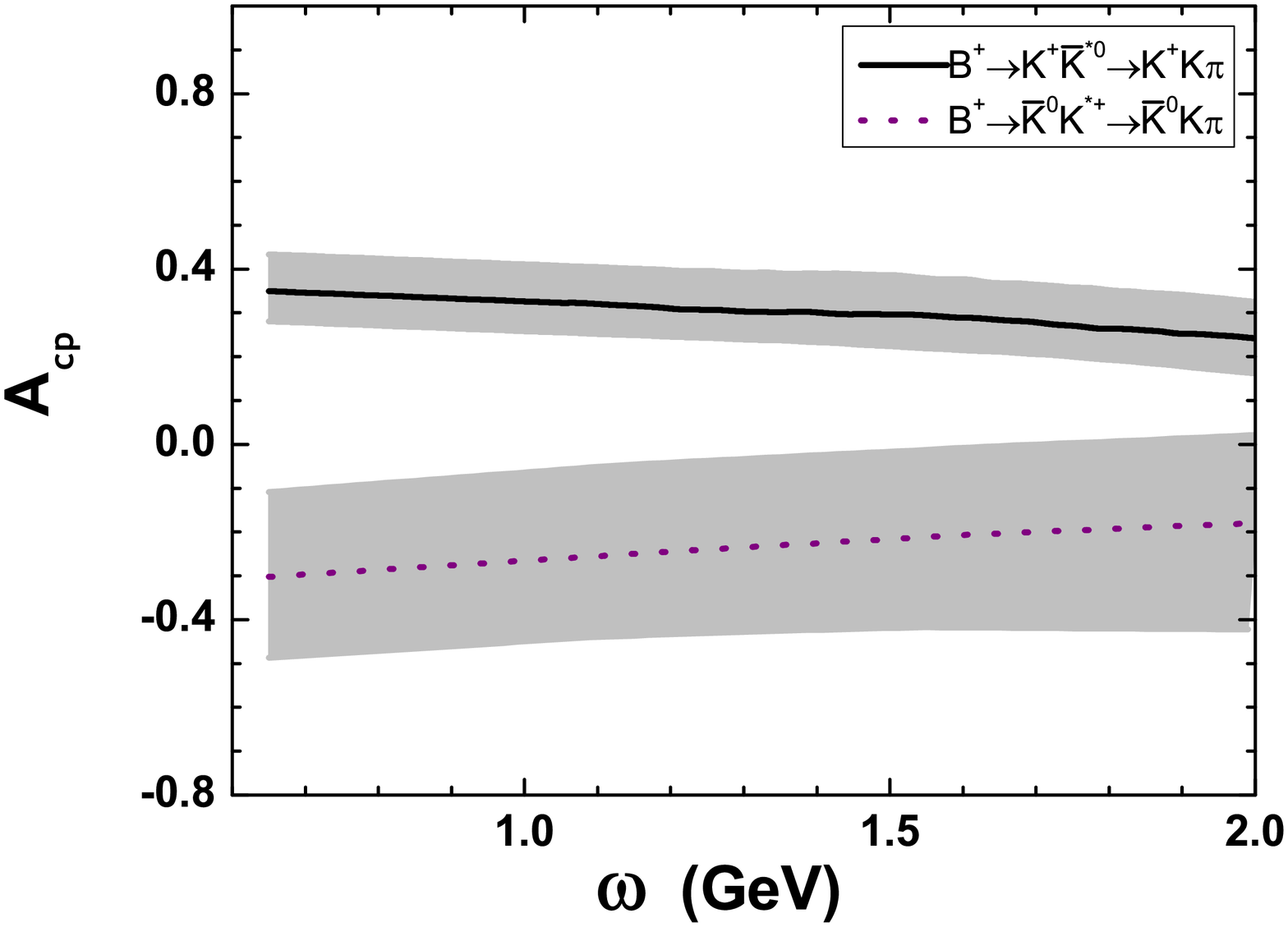}}
\vspace{-0.2cm}
  {\scriptsize\bf (a)\hspace{7.5cm}(b)}
\caption{(a) Differential branching ratios for the $B^+\to K^+\bar{K}^{*0}\to K^+K\pi$ decays,
         and (b) Differential distributions of ${\cal A}_{cp}$ in $\omega$ for the decay modes $B^+\to K^+\bar{K}^{*0}\to K^+K\pi$ and $B^+\to \bar{K}^0 K^{*+}\to \bar{K}^0K\pi$.
         Shaded bands show the estimated theoretical uncertainties.}
\label{fig-br}
\end{figure}

The following input parameters (the masses, decay constants and QCD scale are in units of  GeV) will be used~\cite{pdg2018} in numerical calculations,
\begin{eqnarray}
\Lambda^{(f=4)}_{ \overline{MS} }&=&0.25, \quad m_{B^0}=5.280, \quad m_{B_s}=5.367,
\quad m_{B^\pm}=5.279, \nonumber\\
m_{\pi^{\pm}}&=&0.140, \quad m_{\pi^0}=0.135, \quad
m_{K^{\pm}}=0.494, \quad m_{K^0}=0.498,\nonumber\\
m_{K^{*0}}&=&0.89581, \quad m_{K^{*\pm}}=0.89166,\quad
m_{b}(\text{pole})=4.8, \quad \bar{m_c}(\bar{m_c})=1.275, \nonumber\\
m_s(1{\rm GeV})&=&0.119, \quad \Gamma_{K^*}=0.050, \quad
f_B= 0.19\pm 0.02, \quad f_{B_s}=0.236\pm0.02,\nonumber\\
\tau_{B^0}&=&1.519\; ps, \quad \tau_{B_{s}}=1.512\; ps, \quad \tau_{B^\pm}=1.638\;ps. \label{eq:inputs}
\end{eqnarray}
The $b$-quark mass is chosen as pole mass and the $c$-quark mass corresponds to the running mass in the modified minimal substraction scheme ($\overline{MS}$ scheme),
while the $s$-quark mass is the estimation of the so-called ``current-quark masses''  in a mass-independent subtraction scheme such as $\overline{MS}$ at a scale $\mu \approx 1 {\rm GeV}$.
The values of the Wolfenstein parameters are adopted as given in the Ref.~\cite{pdg2018}:
$A=0.836\pm0.015, \lambda=0.22453\pm 0.00044$, $\bar{\rho} = 0.122^{+0.018}_{-0.017}$, $\bar{\eta}= 0.355^{+0.012}_{-0.011}$.
While the $B$ meson and kaon (pion) DAs are the same as widely adopted in the PQCD approach~\cite{prd95-056008}.

For the decay $B \to  K^*(892) h\to  K \pi h$, the differential branching ratio is described as~\cite{pdg2018},
\begin{eqnarray}
\frac{d{\cal B}}{ds}=\tau_{B}\frac{|\overrightarrow{p_1}||\overrightarrow{p_3}|}{64\pi^3m^3_{B}}|{\cal A}|^2, \label{expr-br}
\end{eqnarray}
with $\tau_{B}$ the mean lifetime of $B$-meson, and  $s=\omega^2$ the invariant mass squared.
The kinematic variables $|\overrightarrow{p_1}|$ and $|\overrightarrow{p_3}|$ denote the kaon momentum and $h$'s momentum in the center-of-mass frame of the $K$-$\pi$ pair,
\begin{eqnarray}
|\overrightarrow{p_1}|=\frac{\sqrt{\lambda(\omega^2,m_K^2,m_{\pi}^2)}}{2\omega}, ~\quad
|\overrightarrow{p_3}|=\frac{\sqrt{\lambda(m_B^2,m_P^2,\omega^2)}}{2\omega}.
\end{eqnarray}

By using the differential branching fraction in Eq.~(\ref{expr-br}), and the decay amplitudes in the Appendix, we calculate
the $CP$ averaged branching ratios ($\cal B$) and the direct $CP$-violating asymmetries ($\cal A_{CP}$) for the concerned decays
$B \to K^*h\to K\pi h$, which are shown in the Table~\ref{kresults} and Table~\ref{piresults} together with some currently available
experimental measurements.
The Gegenbauer moments $a_{1K^*}^{||}=0.05\pm0.02, a_{2K^*}^{||}=0.15\pm0.05$ are determined to cater to the data~\cite{pdg2018,HF2016}, which differ from those in the DAs for a longitudinally polarized $K^*$ meson~\cite{prd76-074018}.
The first theoretical error from the variation of the hard scale $t$ from $0.75t$ to $1.25t$ (without changing $1/b_i$) and the QCD scale $\Lambda_{\rm QCD}=0.25\pm0.05$~GeV, which characterizes the effect of the NLO QCD contributions.
The second error comes from the variations of the shape parameter of the $B_{(s)}$ meson distribution amplitude
$\omega_B=0.40\pm0.04$~GeV or $\omega_{B_s}=0.50\pm0.05$~GeV~\cite{prd63-054008,epjc23-275,prd63-074009,plb504-6}.
The last one is caused by the Gegenbauer moments $a_{1K^*}^{||}=0.05\pm0.02, a_{2K^*}^{||}=0.15\pm0.05$.
The first two errors are comparable and contribute the main uncertainties in our approach, while the last one is less than $15\%$.
The errors from $\tau_{B^\pm}$, $\tau_{B^0}$, $\tau_{B_s}$ and the Wolfenstein parameters in~\cite{pdg2018} are very small and have been neglected.

From the numerical results as shown in those two tables, one can address some issues as follows:
\begin{itemize}  
\item[(1)]
The isospin conservation is assumed for the strong decays of an $I=1/2$ resonance $K^*$ to $K\pi$ when we compute the branching fractions of the quasi-two-body process $B \to K^*h\to K\pi h$, namely,
\begin{eqnarray}
\frac{\Gamma(K^{*0} \to K^+\pi^-)}{\Gamma(K^{*0} \to K\pi)}=2/3, ~\quad
\frac{\Gamma(K^{*+} \to K^+\pi^0)}{\Gamma(K^{*+} \to K\pi)}=1/3.
\end{eqnarray}
Taking $B^0 \to \pi^0(K^{*0}\to)K \pi$ decay as an example, we can obtain the quasi-two-body branching fraction ${\cal B}(B^0 \to \pi^0(K^{*0}\to)K^+ \pi^-)$
under the narrow width approximation relation
\begin{eqnarray}
\mathcal{B}(B^0 \to \pi^0 K^{*0}\to \pi^0 K \pi ) &=&
\mathcal{B}( B^0 \to \pi^0 K^{*0}) \cdot {\mathcal B}(K^{*0} \to K\pi),\label{eq:def1}\nonumber\\
\mathcal{B}(B^0 \to \pi^0 K^{*0}\to \pi^0 K^+ \pi^- ) &=&
\mathcal{B}( B^0 \to \pi^0 K^{*0}) \cdot {\mathcal B}(K^{*0} \to K^+\pi^-), \label{eq:def2}
\end{eqnarray}
where we assume the $K^*\to K\pi$ branching fraction to be 100\%.

\item[(2)]
It is worth of stressing that there already exist many well known results for $B_{(s)} \to hK^*$ in the two-body framework both in the PQCD~\cite{prd74-094020,prd76-074018,epjc59-49,npb931-79} and QCDF~\cite{npb675-333,prd80-114008,prd80-114026} approaches.
In the narrow width limit, the branching ratios of the two-body decays $B_{(s)} \to hK^*$ are extracted from the corresponding  quasi-two-body decay modes as listed in Table~\ref{kresults} and Table~\ref{piresults}.
One can see that the branching ratios of the quasi-two-body decay modes are in good agreement with those two-body analyses as presented in Refs.~\cite{npb931-79,prd74-094020,prd76-074018,epjc59-49} in PQCD approach.
The consistency between the theoretical predictions and the measured values for the branching ratios supports the PQCD factorization for exclusive hadronic $B$ meson decays.
The measured $CP$ violation is just a number in two-body $B$ decays, while in three-body decays, one can measure the distribution of $CP$ asymmetry.
The $CP$ asymmetry in the three-body framework is moderated by the finite width of the $K^*$ resonance appearing in the time-like form factor $F_{K\pi}$.
It may be more appropriate to treat $B\to hK^*$ as three-body decays.
By comparing with corresponding results in the QCDF approach~\cite{npb675-333,prd80-114008,prd80-114026}, we find that the PQCD predictions for the branching ratios as listed in Table~\ref{kresults} and ~\ref{piresults} are similar to the QCDF results.
Since the mechanism and the source of the $CP$ asymmetries for the considered decay modes are very different in the PQCD approach and the QCDF approach,
the QCDF results for the direct $CP$ asymmetries are quite different from ours.
Because currently available experimental measurements still have relatively large uncertainties, we have to wait for more time to test these different predictions.

\item[(3)]
We calculated the branching ratios and $CP$ violations of the quasi-two-body decays $B \to K K^* \to K K\pi$ as shown in Table~\ref{kresults}.
Moreover, there is no $CP$ violation for the decays $B^0 \to K^0(\bar{K}^{*0}\to)K \pi$, $B^0 \to \bar{K}^0(K^{*0}\to)K \pi$, $B_s^0 \to K^0(\bar{K}^{*0}\to)K \pi$ and $B_s^0 \to \bar{K}^0(K^{*0}\to)K \pi$ within the standard model, since there is only one kind of penguin
operator involved in the decay amplitudes of the considered decays, which can be seen from Eqs.~(\ref{amp7}-\ref{amp10}).
The PQCD predictions of the sum of branching ratios of $B_s^0 \to K^+(K^{*-}\to)K\pi$ and $B_s^0 \to K^-(K^{*+}\to)K \pi$ decays, as well as the sum of branching ratios of $B_s^0 \to K^0(\bar{K}^{*0}\to)K \pi$ and $B_s^0 \to \bar{K}^0(K^{*0}\to)K \pi$ are in consistent with the LHCb measurements~\cite{NJP16-123001,jhep01-012} and support their first observations of $B_s^0$ meson decays to $K^{*\pm}K^\mp$ and $K^0_SK^{*0}$.
The LHCb Collaboration reported that there is no evidence for the decay $B^0 \to K^0_SK^{*0}$ and an upper limit is set on the branching ratio.
Our result for the ${\cal B}(B^0 \to K^0(\bar{K}^{*0}\to)K \pi)$ plus ${\cal B}(B^0 \to \bar{K}^0(K^{*0}\to)K \pi)$ is around $0.54 \times 10^{-6}$, which can be examined in the forthcoming experiments.

\item[(4)]
For the considered decay channels $B \to \pi (K^* \to) K \pi$, there are already some experimental measurements for the branching ratios and $CP$ asymmetries shown in the forth column in Table~\ref{piresults}.
Although the error bars of the $CP$ violations are still large, one can find that our theoretical calculations have the same sign as these measured entries.
For the four decay modes, ${\cal B}( B^+ \to \pi^+(K^{*0}\to)K \pi)=(7.12^{+2.77}_{-2.07})\times10^{-6}$, ${\cal B}(B^0 \to \pi^-(K^{*+}\to)K \pi)=(6.51^{+2.33}_{-1.75})\times10^{-6}$, ${\cal B}(B^+ \to \pi^0(K^{*+}\to)K \pi)=(5.00^{+1.78}_{-1.34})\times 10^{-6}$ and ${\cal B}(B^0 \to \pi^0(K^{*0}\to)K \pi)
=(2.07^{+0.83}_{-0.62})\times10^{-6}$, the PQCD predictions are in agreement with the world averages within errors.
When more data become available, we do recommend the LHCb and/or Belle-II experiments to remeasure the direct $CP$ asymmetry
in channels like $B^+ \to \pi^0(K^{*+}\to)K \pi$, $B^0 \to \pi^0(K^{*0}\to)K \pi$ and so on,
because these decay modes may have large branching ratios and large direct $CP$ asymmetries.

\item[(5)]
For the $B_s^0 \to \pi^+(K^{*-}\to)K\pi$ decay process, our prediction is ${\cal B}=(8.08^{+2.88}_{-2.09}) \times 10^{-6}$ at leading-order in the quasi-two-body framework in this work, such a branching ratio is a bit larger than the value $(3.3\pm 1.2)\times 10^{-6}$ in~\cite{pdg2018}.
For the corresponding two-body modes, the previous theoretical predictions as given in Refs.~\cite{npb675-333,prd76-074018,prd80-114026} are larger than the data as well.
In Ref.~\cite{npb931-79}, the authors considered the next-to-leading-order (NLO) corrections and found that the NLO contribution will result in a 37\% reduction of the leading order PQCD prediction for the ``tree'' dominated decay $B_s^0 \to \pi^+K^{*-}$.
The authors confirmed that the branching ratios of the quasi-two-body modes in the three-body and two-body frameworks are close to each other in Ref.~\cite{plb763-29}, since the ${\cal B}(\rho \to \pi\pi)\approx 100\%$.
Compared with the previous calculations of the two-body decays $B_{(s)} \to P K^*$ from PQCD~\cite{prd74-094020,prd76-074018,epjc59-49,npb931-79}, we can obtain the consistency between the two-body and three-body modes.
Maybe we can assume that the PQCD prediction of the branching ratio of the quasi-to-body decay $B_s^0 \to \pi^+(K^{*-}\to)K\pi$ will accommodate to data if we take the NLO contributions into consideration in the three-body framework.
However, how to evaluate the NLO corrections to the three-body decays in the PQCD framework is a big task and will be left for the future studies.
\end{itemize} 

In Fig.~\ref{fig-br}(a), we show the $\omega$-dependence of differential decay rate
$d{\cal B}(B^+ \to K^+\bar{K}^{*0} \to K^+K\pi)/d\omega$.
The $\bar{K}^{*0}$ is visible as a narrow peak near $0.89$ GeV.
We find that the main portion of the branching ratios lies in the region around the pole mass of the $K^*$ resonance as expected by examining the distribution of the branching ratios in the kaon-pion invariant mass $\omega$.
The central values of ${\cal B}$ are $0.25\times 10^{-7}$ and $0.37\times 10^{-7}$ when the integration over $\omega$ is limited in the range of $\omega=[m_{K^*}-0.5\Gamma_{K^*}, m_{K^*}+0.5\Gamma_{K^*}]$ or
$\omega=[m_{K^*}-\Gamma_{K^*}, m_{K^*}+\Gamma_{K^*}]$ respectively, which amount to
$50\%$ and $74\%$ of the total branching ratio ${\cal B}=0.50\times10^{-7}$ as listed in Table~\ref{kresults}.
In two-body $B$ decays, the measured $CP$ violation is just a number due to the fixed kinematics.
While in three-body decays, the decay amplitudes depend on the $K\pi$ invariant mass, which resulting in the differential distribution of direct $CP$ asymmetries.
In Fig.~\ref{fig-br}(b), we display the differential distributions of ${\mathcal A}_{CP}$ for the two decay modes $B^+\to K^+\bar{K}^{*0}\to K^+K\pi$ (black solid line) and $B^+\to \bar{K}^0 K^{*+}\to \bar{K}^0K\pi$ (purple dotted line), respectively.
One can find a falloff of ${\mathcal A}_{CP}$ with $\omega$ for
$B^+\to K^+\bar{K}^{*0}\to K^+K\pi$.
It implies that the direct $CP$ asymmetries in the above three quasi-two-body
decays, if calculated as the two-body decays with the $K^*$ resonance mass being fixed to $m_{K^*}$, may be overestimated.
The ascent of the differential distribution of ${\mathcal A}_{CP}$ with $\omega$
for $B^+\to \bar{K}^0 K^{*+}\to \bar{K}^0K\pi$ suggests that its direct $CP$ asymmetry,
if calculated in the two-body formalism, may be underestimated.

\begin{table}[thp]
\caption{The $CP$ averaged branching ratios and direct $CP$-violating asymmetries
of $B_{(s)}\to K (K^* \to) K \pi$ decays calculated in PQCD approach together
with experimental data~\cite{pdg2018,HF2016}.
The theoretical errors corresponding to the uncertainties due to the next-to-leading-order effects (the hard scale $t$ and the QCD scale $\Lambda_{\rm QCD}$), the shape parameters $\omega_{B_{(s)}}$ in the wave function of $B_{(s)}$ meson and the Gegenbauer moments ($a_{1K^*}^{||}$ and $a_{2K^*}^{||}$), respectively.}
\label{kresults}
\begin{center}
\begin{tabular}{cccc}
 \hline \hline
{Modes}          &\qquad  & Quasi-two-body results              &  Experiment   \\
\hline
 $B^+ \to K^+(\bar{K}^{*0} \to)K\pi$         &~~~${\cal B} (10^{-6})$~~~     &$0.50^{+0.13+0.13+0.02}_{-0.10-0.10-0.02}$  &$~~<1.1~~$\\
                                 &$\cal A_{CP} (\%)$    &$35.0^{+4.6+1.7+1.4}_{-4.4-1.4-1.6}$                   &$-$\\
  $B^0 \to K^+(K^{*-} \to)K\pi$       &~~~${\cal B} (10^{-6})$~~~     &$0.06^{+0.01+0.00+0.00}_{-0.00-0.01-0.01}$    &$~~<0.4\footnotemark[1]~~$\\
                                 &$\cal A_{CP} (\%)$    &$39.1^{+4.6+2.8+6.4}_{-9.6-0.0-3.4}$                    &$-$\\
  $B^0 \to K^-(K^{*+} \to)K\pi$      &~~~${\cal B} (10^{-6})$~~~     &$0.07^{+0.02+0.01+0.01}_{-0.00-0.00-0.00}$    &$~~<0.4\footnotemark[1]~~$\\
                                 &$\cal A_{CP} (\%)$    &$28.8^{+0.0+3.2+5.1}_{-7.4-0.0-2.1}$                 &$-$\\
  $B_s^0 \to K^+(K^{*-}\to)K\pi$        &~~~${\cal B} (10^{-6})$~~~     &$7.27^{+1.55+0.45+0.81}_{-1.66-0.37-0.77}$   &$~~(12.5\pm2.6)\footnotemark[1]~~$\\
                                 &$\cal A_{CP} (\%)$    &$60.0^{+5.5+7.3+2.1}_{-5.4-6.3-2.1}$                  &$-$\\

  $B_s^0 \to K^-(K^{*+}\to)K \pi$       &~~~${\cal B} (10^{-6})$~~~     &$6.96^{+2.27+1.64+0.32}_{-1.60-1.10-0.31}$    &$~~(12.5\pm2.6)\footnotemark[1]~~$\\
                                 &$\cal A_{CP} (\%)$    &$-46.3^{+5.8+3.0+1.5}_{-5.8-1.8-1.4}$                 &$-$\\
  $B^+ \to \bar{K}^0(K^{*+}\to)K \pi$  &~~~${\cal B} (10^{-6})$~~~     &$0.12^{+0.02+0.00+0.01}_{-0.05-0.01-0.01}$    &$-$\\
                                 &$\cal A_{CP} (\%)$    &$-12.5^{+11.5+6.1+1.5}_{-12.7-1.2-1.4}$                 &$-$\\
  $B^0 \to K^0(\bar{K}^{*0}\to)K \pi$        &~~~${\cal B} (10^{-6})$~~~     &$0.40^{+0.11+0.12+0.00}_{-0.09-0.09-0.01}$    &$<0.96\footnotemark[1]$\\
                                 &$\cal A_{CP} (\%)$    &$0$                  &$-$\\
  $B^0 \to \bar{K}^0(K^{*0}\to)K \pi$         &~~~${\cal B} (10^{-6})$~~~     &$0.14^{+0.03+0.01+0.03}_{-0.03-0.00-0.02}$    &$<0.96\footnotemark[1]$\\
                                 &$\cal A_{CP} (\%)$    &$0$                 &$-$\\
  $B_s^0 \to K^0(\bar{K}^{*0}\to)K \pi$         &~~~${\cal B} (10^{-6})$~~~     &$6.19^{+1.45+0.12+0.81}_{-1.56-0.14-0.77}$    &$(16.4\pm4.1)\footnotemark[1]$\\
                                 &$\cal A_{CP} (\%)$   &$0$                  &$-$\\
  $B_s^0 \to \bar{K}^0(K^{*0}\to)K \pi$       &~~~${\cal B} (10^{-6})$~~~     &$7.16^{+2.55+1.78+0.31}_{-1.84-1.17-0.28}$   &$(16.4\pm4.1)\footnotemark[1]$\\
                                 &$\cal A_{CP} (\%)$    &$0$                  &$-$\\

 \hline \hline
\end{tabular}
\end{center}
\footnotetext[1]{Includes two distinct decay processes: ${\cal B}(B_{(s)}\to f) + {\cal B}(B_{(s)}\to \bar{f})$.}
\end{table}

\begin{table}[thb]
\caption{The $CP$ averaged branching ratios and direct $CP$-violating asymmetries
of $B_{(s)}\to \pi (K^* \to) K \pi$ decays calculated in PQCD approach together
with experimental data~\cite{pdg2018,HF2016}.
The theoretical errors corresponding to the uncertainties due to the next-to-leading-order effects (the hard scale $t$ and the QCD scale $\Lambda_{\rm QCD}$), the shape parameters $\omega_{B_{(s)}}$ in the wave function of $B_{(s)}$ meson and the Gegenbauer moments ($a_{1K^*}^{||}$ and $a_{2K^*}^{||}$), respectively.}
 \label{piresults}
\begin{center}
\begin{tabular}{cccc}
 \hline \hline
{Modes}          &\qquad  & Quasi-two-body results              &  Experiment   \\
\hline
$B^+ \to \pi^+(K^{*0}\to)K \pi$     &~~~${\cal B} (10^{-6})$~~~     &$7.12^{+1.73+2.16+0.16}_{-1.44-1.49-0.15}$    &$~~10.1\pm{0.9}~~$\\
                                 &$\cal A_{CP} (\%)$    &$-2.5^{+1.6+0.4+0.1}_{-0.3-0.0-0.1}$                &$-$\\
  $B^0 \to \pi^-(K^{*+}\to)K \pi$         &~~~${\cal B} (10^{-6})$~~~     &$6.51^{+1.42+1.84+0.20}_{-1.19-1.27-0.17}$    &$~~8.4\pm{0.8}~~$\\
                                 &$\cal A_{CP} (\%)$    &$-47.0^{+3.7+4.7+1.4}_{-2.2-4.4-1.5}$               &$-22\pm{6}$\\
  $B_s^0 \to \pi^+(K^{*-}\to)K\pi$       &~~~${\cal B} (10^{-6})$~~~     &$8.08^{+0.37+2.83+0.36}_{-0.49-2.00-0.35}$    &$~~3.3\pm{1.2}~~$\\
                                 &$\cal A_{CP} (\%)$   &$-21.9^{+2.3+2.7+0.7}_{-2.3-3.0-0.9}$                   &$-$\\
  $B^+ \to \pi^0(K^{*+}\to)K \pi$      &~~~${\cal B} (10^{-6})$~~~     &$5.00^{+0.95+1.50+0.13}_{-0.81-1.06-0.11}$    &$~~8.2\pm{1.9}~~$\\
                                 &$\cal A_{CP} (\%)$    &$-30.5^{+2.7+4.5+1.2}_{-1.0-4.3-1.1}$               &$-6\pm{24}$\\
  $B^0 \to \pi^0(K^{*0}\to)K \pi$       &~~~${\cal B} (10^{-6})$~~~     &$2.07^{+0.62+0.55+0.07}_{-0.50-0.37-0.06}$    &$~~3.3\pm{0.6}~~$\\
                                 &$\cal A_{CP} (\%)$    &$-6.9^{+2.6+1.1+0.2}_{-0.1-0.7-0.2}$                 &$-15\pm{13}$\\
  $B_s^0 \to \pi^0(\bar{K}^{*0}\to)K \pi$        &~~~${\cal B} (10^{-6})$~~~     &$0.09^{+0.05+0.02+0.01}_{-0.02-0.01-0.01}$   &$-$\\
                                 &$\cal A_{CP} (\%)$    &$-67.4^{+24.3+9.2+1.4}_{-14.9-8.5-1.0}$                  &$-$\\
 \hline \hline
\end{tabular}
\end{center}
\end{table}

\section{Summary}
In this paper, we calculated the quasi-two-body decays $B_{(s)}\to K^*(892)h\to K\pi h$ by using the PQCD factorization approach.
The relativistic Breit-Wigner formula for the $P$-wave narrow resonance $K^*(892)$ was adopted to parameterize the time-like form factor $F_{K\pi}$. The kaon-pion distribution amplitude $\Phi_{K\pi}^{\rm P}$ with the $P$-wave time-like form factor $F_{K\pi}$ was
employed to describe the resonant state $K^*$ and its interactions with the kaon-pion pair. We predicted the branching ratios and the
direct $CP$ asymmetries of the concerned decay channels, and compared their differential branching ratios with currently available
data. General agreements between the PQCD predictions and the data can be achieved by tuning the Gegenbauer moments of the
$P$-wave kaon-pion DAs. The direct $CP$ asymmetry of the $B_{(s)}\to K^*(892)h\to K\pi h$ modes are not numbers but depend on
the kaon-pion invariant mass. Owing to the isospin conservation in $K^*(892) \to K\pi$ decays, we can obtain the separate branching
ratios of the corresponding quasi-two-body decays. More precise data from the LHCb and the future Belle II will test our predictions.

\begin{acknowledgments}
Many thanks to Hsiang-nan Li and Rui Zhou for valuable discussions.
This work was supported by the National Natural Science Foundation of China under the Grant No.~11775117 , 11547038 and 11235005.
\end{acknowledgments}

\appendix
\section{Decay amplitudes}
The decay amplitudes for considered quasi-two-body decay modes in this work are given as follows:

\begin{eqnarray}
{\cal A}(B^+ \to K^+(\bar{K}^{*0} \to)K\pi) &=& \frac{G_F} {\sqrt{2}} \big\{V_{ub}^*V_{ud}[(\frac{C_1}{3}+C_2)F^{LL}_{aP}+C_1M^{LL}_{aP}]-V_{tb}^*V_{td}[(\frac{C_3}{3}+C_4-\frac{C_9}{6}-\frac{C_{10}}{2})F^{LL}_{eP}\nonumber\\
&+&(C_3-\frac{C_9}{2})M^{LL}_{eP}+(C_5-\frac{C_7}{2})M^{LR}_{eP}+(\frac{C_3}{3}+C_4+\frac{C_9}{3}+C_{10})F^{LL}_{aP}\nonumber\\
&+&(\frac{C_5}{3}+C_6+\frac{C_7}{3}+C_8)F^{SP}_{aP}+(C_3+C_9)M^{LL}_{aP}+(C_5+C_7)M^{LR}_{aP}]\big\} \;, \label{amp1}
\end{eqnarray}
 \begin{eqnarray}
{\cal A}(B^0 \to K^+(K^{*-} \to)K\pi) &=& \frac{G_F} {\sqrt{2}} \big\{V_{ub}^*V_{ud}[(C_1+\frac{C_2}{3})F^{LL}_{aP}+C_2M^{LL}_{aP}]-V_{tb}^*V_{td}[(C_3+\frac{C_4}{3}-\frac{C_9}{2}-\frac{C_{10}}{6}\nonumber\\
&-&C_5-\frac{C_6}{3}+\frac{C_7}{2}+\frac{C_8}{6})F^{LL}_{aK^*}+(C_4-\frac{C_{10}}{2})M^{LL}_{aK^*}+(C_6-\frac{C_8}{2})M^{SP}_{aK^*}\nonumber\\
&+&(C_3+\frac{C_4}{3}+C_9+\frac{C_{10}}{3}-C_5-\frac{C_6}{3}-C_7-\frac{C_8}{3})F^{LL}_{aP}\nonumber\\
&+&(C_4+C_{10})M^{LL}_{aP}+(C_6+C_8)M^{SP}_{aP}]\big\} \;,
\label{amp2}
\end{eqnarray}
\begin{eqnarray}
{\cal A}(B^0 \to K^-(K^{*+} \to)K\pi) &=& \frac{G_F} {\sqrt{2}} \big\{V_{ub}^*V_{ud}[(C_1+\frac{C_2}{3})F^{LL}_{aK^*}+C_2 M^{LL}_{aK^*}]-V_{tb}^*V_{td}[(C_3+\frac{C_4}{3}+C_9+\frac{C_{10}}{3}\nonumber\\
&-&C_5-\frac{C_6}{3}-C_7-\frac{C_8}{3})F^{LL}_{aK^*}+(C_4+C_{10})M^{LL}_{aK^*}+(C_6+C_8)M^{SP}_{aK^*}\nonumber\\
&+&(C_3+\frac{C_4}{3}-\frac{C_9}{2}-\frac{C_{10}}{6}-C_5-\frac{C_6}{3}+\frac{C_7}{2}+\frac{C_8}{6})F^{LL}_{aP}\nonumber\\
&+&(C_4-\frac{C_{10}}{2})M^{LL}_{aP}+(C_6-\frac{C_8}{2})M^{SP}_{aP}]\big\} \;,\nonumber\\
\label{amp3}
\end{eqnarray}
\begin{eqnarray}
 {\cal A}(B_s^0 \to K^+(K^{*-}\to)K\pi) &=&  \frac{G_F} {\sqrt{2}} \big\{V_{ub}^*V_{us}[(\frac{C_1}{3}+C_2)F^{LL}_{eK^*}+C_1 M^{LL}_{eK^*}+(C_1+\frac{C_2}{3})F^{LL}_{aP}+C_2M^{LL}_{aP}]\nonumber\\
&-&V_{tb}^*V_{ts}[(\frac{C_3}{3}+C_4+\frac{C_9}{3}+C_{10})F^{LL}_{eK^*}+(\frac{C_5}{3}+C_6+\frac{C_7}{3}+C_8)F^{SP}_{eK^*}+(C_3+C_9)M^{LL}_{eK^*}\nonumber\\
&+&(C_5+C_7)M^{LR}_{eK^*}+(\frac{4}{3}(C_3+C_4-\frac{C_9}{2}-\frac{C_{10}}{2})-C_5-\frac{C_6}{3}+\frac{C_7}{2}+\frac{C_8}{6})F^{LL}_{aK^*}\nonumber\\
&+&(\frac{C_5}{3}+C_6-\frac{C_7}{6}-\frac{C_8}{2})F^{SP}_{aK^*}+(C_3+C_4-\frac{C_9}{2}-\frac{C_{10}}{2})M^{LL}_{aK^*}+(C_5-\frac{C_7}{2})M^{LR}_{aK^*}\nonumber\\
&+&(C_6-\frac{C_8}{2})M^{SP}_{aK^*}+(C_3+\frac{C_4}{3}+C_9+\frac{C_{10}}{3}-C_5-\frac{C_6}{3}-C_7-\frac{C_8}{3})F^{LL}_{aP}\nonumber\\
&+&(C_4+C_{10})M^{LL}_{aP}+(C_6+C_8)M^{SP}_{aP}]\big\} \;,\label{amp4}
\end{eqnarray}
\begin{eqnarray}
 {\cal A}(B_s^0 \to K^-(K^{*+}\to)K \pi) &=&  \frac{G_F} {\sqrt{2}} \big\{V_{ub}^*V_{us}[(C_1+\frac{C_2}{3})F^{LL}_{aK^*}+C_2 M^{LL}_{aK^*}+(\frac{C_1}{3}+C_2)F^{LL}_{eP}+C_1M^{LL}_{eP}]\nonumber\\
&-&V_{tb}^*V_{ts}[(C_3+\frac{C_4}{3}+C_9+\frac{C_{10}}{3}-C_5-\frac{C_6}{3}-C_7-\frac{C_8}{3})F^{LL}_{aK^*}+(C_4+C_{10})M^{LL}_{aK^*}\nonumber\\
&+&(C_6+C_8)M^{SP}_{aK^*}+(\frac{C_3}{3}+C_4+\frac{C_9}{3}+C_{10})F^{LL}_{eP}+(C_3+C_9)M^{LL}_{eP}+(C_5+C_7)M^{LR}_{eP}\nonumber\\
&+&(\frac{4}{3}(C_3+C_4-\frac{C_9}{2}-\frac{C_{10}}{2})-C_5-\frac{C_6}{3}+\frac{C_7}{2}+\frac{C_8}{6})F^{LL}_{aP}\nonumber\\
&+&(\frac{C_5}{3}+C_6-\frac{C_7}{6}-\frac{C_8}{2})F^{SP}_{aP}+(C_3+C_4-\frac{C_9}{2}-\frac{C_{10}}{2})M^{LL}_{aP}\nonumber\\
&+&(C_5-\frac{C_7}{2})M^{LR}_{aP}+(C_6-\frac{C_8}{2})M^{SP}_{aP}
]\big\} \;,\label{amp5}
\end{eqnarray}
\begin{eqnarray}
 {\cal A}(B^+ \to \bar{K}^0(K^{*+}\to)K \pi) &=& \frac{G_F} {\sqrt{2}}\big\{V_{ub}^*V_{ud}[(\frac{C_1}{3}+C_2)F^{LL}_{aK^*}+C_1 M^{LL}_{aK^*}]-V_{tb}^*V_{td}[(\frac{C_3}{3}+C_4-\frac{C_9}{6}-\frac{C_{10}}{2})F^{LL}_{eK^*}\nonumber\\
&+&(\frac{C_5}{3}+C_6-\frac{C_7}{6}-\frac{C_8}{2})F^{SP}_{eK^*}+(C_3-\frac{C_9}{2})M^{LL}_{eK^*}+(C_5-\frac{C_7}{2})M^{LR}_{eK^*}\nonumber\\
&+&(\frac{C_3}{3}+C_4+\frac{C_9}{3}+C_{10})F^{LL}_{aK^*}+(\frac{C_5}{3}+C_6+\frac{C_7}{3}+C_8)F^{SP}_{aK^*}\nonumber\\
&+&(C_3+C_9)M^{LL}_{aK^*}+(C_5+C_7)M^{LR}_{aK^*}]\big\} \;,\label{amp6}
\end{eqnarray}
\begin{eqnarray}
{\cal A}(B^0 \to K^0(\bar{K}^{*0}\to)K \pi) &=& -\frac{G_F} {\sqrt{2}}\big\{V_{tb}^*V_{td}[(C_3+\frac{C_4}{3}-\frac{C_9}{2}-\frac{C_{10}}{6}-C_5-\frac{C_6}{3}+\frac{C_7}{2}+\frac{C_8}{6})F^{LL}_{aK^*}\nonumber\\
&+&(C_4-\frac{C_{10}}{2})M^{LL}_{aK^*}+(C_6-\frac{C_8}{2})(M^{SP}_{aK^*}+M^{SP}_{aP})+(\frac{C_3}{3}+C_4-\frac{C_9}{6}-\frac{C_{10}}{2})F^{LL}_{eP}\nonumber\\
&+&(C_3-\frac{C_9}{2})M^{LL}_{eP}+(C_5-\frac{C_7}{2})(M^{LR}_{eP}+M^{LR}_{aP})+(\frac{C_5}{3}+C_6-\frac{C_7}{6}-\frac{C_8}{2})F^{SP}_{aP}\nonumber\\
&+&(\frac{4}{3}(C_3+C_4-\frac{C_9}{2}-\frac{C_{10}}{2})-C_5-\frac{C_6}{3}+\frac{C_7}{2}+\frac{C_8}{6})F^{LL}_{aP}+(C_3+C_4-\frac{C_9}{2}-\frac{C_{10}}{2})M^{LL}_{aP}]\big\} \;,\nonumber\\
\label{amp7}
 \end{eqnarray}
 \begin{eqnarray}
{\cal A}(B^0 \to \bar{K}^0(K^{*0}\to)K \pi) &=& -\frac{G_F} {\sqrt{2}}\big\{V_{tb}^*V_{td}[(\frac{C_3}{3}+C_4-\frac{C_9}{6}-\frac{C_{10}}{2})F^{LL}_{eK^*}+(\frac{C_5}{3}+C_6-\frac{C_7}{6}-\frac{C_8}{2})(F^{SP}_{eK^*}+F^{SP}_{aK^*})\nonumber\\
&+&(C_3-\frac{C_9}{2})M^{LL}_{eK^*}
+(\frac{4}{3}(C_3+C_4-\frac{C_9}{2}-\frac{C_{10}}{2})-C_5-\frac{C_6}{3}+\frac{C_7}{2}+\frac{C_8}{6})F^{LL}_{aK^*}\nonumber\\
&+&(C_3+C_4-\frac{C_9}{2}-\frac{C_{10}}{2})M^{LL}_{aK^*}+(C_5-\frac{C_7}{2})(M^{LR}_{eK^*}+M^{LR}_{aK^*})+(C_6-\frac{C_8}{2})(M^{SP}_{aK^*}\nonumber\\
&+&M^{SP}_{aP})+(C_3+\frac{C_4}{3}-\frac{C_9}{2}-\frac{C_{10}}{6}-C_5-\frac{C_6}{3}+\frac{C_7}{2}+\frac{C_8}{6})F^{LL}_{aP}+(C_4-\frac{C_{10}}{2})M^{LL}_{aP}]\big\} \;,\nonumber\\
\label{amp8}
 \end{eqnarray}
 \begin{eqnarray}
{\cal A}(B_s^0 \to K^0(\bar{K}^{*0}\to)K \pi) &=& -\frac{G_F} {\sqrt{2}}\big\{V_{tb}^*V_{ts}[(\frac{C_3}{3}+C_4-\frac{C_9}{6}-\frac{C_{10}}{2})F^{LL}_{eK^*}+(\frac{C_5}{3}+C_6-\frac{C_7}{6}-\frac{C_8}{2})(F^{SP}_{eK^*}+F^{SP}_{aK^*})\nonumber\\
&+&(C_3-\frac{C_9}{2})M^{LL}_{eK^*}
+(\frac{4}{3}(C_3+C_4-\frac{C_9}{2}-\frac{C_{10}}{2})-C_5-\frac{C_6}{3}+\frac{C_7}{2}+\frac{C_8}{6})F^{LL}_{aK^*}\nonumber\\
&+&(C_3+C_4-\frac{C_9}{2}-\frac{C_{10}}{2})M^{LL}_{aK^*}+(C_5-\frac{C_7}{2})(M^{LR}_{eK^*}+M^{LR}_{aK^*})+(C_6-\frac{C_8}{2})(M^{SP}_{aK^*}\nonumber\\
&+&M^{SP}_{aP})+(C_3+\frac{C_4}{3}-\frac{C_9}{2}-\frac{C_{10}}{6}-C_5-\frac{C_6}{3}+\frac{C_7}{2}+\frac{C_8}{6})F^{LL}_{aP}+(C_4-\frac{C_{10}}{2})M^{LL}_{aP}]\big\}  \;,\nonumber\\
\label{amp9}
\end{eqnarray}
\begin{eqnarray}
{\cal A}(B_s^0 \to \bar{K}^0(K^{*0}\to)K \pi) &=& -\frac{G_F} {\sqrt{2}}\big\{V_{tb}^*V_{ts}[(C_3+\frac{C_4}{3}-\frac{C_9}{2}-\frac{C_{10}}{6}-C_5-\frac{C_6}{3}+\frac{C_7}{2}+\frac{C_8}{6})F^{LL}_{aK^*}\nonumber\\
&+&(C_4-\frac{C_{10}}{2})M^{LL}_{aK^*}+(C_6-\frac{C_8}{2})(M^{SP}_{aK^*}+M^{SP}_{aP})+(\frac{C_3}{3}+C_4-\frac{C_9}{6}-\frac{C_{10}}{2})F^{LL}_{eP}\nonumber\\
&+&(C_3-\frac{C_9}{2})M^{LL}_{eP}+(C_5-\frac{C_7}{2})(M^{LR}_{eP}+M^{LR}_{aP})+(\frac{C_5}{3}+C_6-\frac{C_7}{6}-\frac{C_8}{2})F^{SP}_{aP}\nonumber\\
&+&(\frac{4}{3}(C_3+C_4-\frac{C_9}{2}-\frac{C_{10}}{2})-C_5-\frac{C_6}{3}+\frac{C_7}{2}+\frac{C_8}{6})F^{LL}_{aP}+(C_3+C_4-\frac{C_9}{2}-\frac{C_{10}}{2})M^{LL}_{aP}]\big\} \;,\nonumber\\
\label{amp10}
\end{eqnarray}
 \begin{eqnarray}
{\cal A}(B^+ \to \pi^+(K^{*0}\to)K \pi) &=& \frac{G_F} {\sqrt{2}}\big\{V_{ub}^*V_{us}[(\frac{C_1}{3}+C_2)F^{LL}_{aP}+C_1M^{LL}_{aP}]-V_{tb}^*V_{ts}[(\frac{C_3}{3}+C_4-\frac{C_9}{6}-\frac{C_{10}}{2})F^{LL}_{eP}\nonumber\\
&+&(C_3-\frac{C_9}{2})M^{LL}_{eP}+(C_5-\frac{C_7}{2})M^{LR}_{eP}+(\frac{C_3}{3}+C_4+\frac{C_9}{3}+C_{10})F^{LL}_{aP}\nonumber\\
&+&(\frac{C_5}{3}+C_6+\frac{C_7}{3}+C_8)F^{SP}_{aP}+(C_3+C_9)M^{LL}_{aP}+(C_5+C_7)M^{LR}_{aP}]\big\} \;,\nonumber\\
 \label{amp11}
\end{eqnarray}
 \begin{eqnarray}
 {\cal A}(B^0 \to \pi^-(K^{*+}\to)K \pi) &=& \frac{G_F} {\sqrt{2}}\big\{V_{ub}^*V_{us}[(\frac{C_1}{3}+C_2)F^{LL}_{eP}+C_1M^{LL}_{eP}]-V_{tb}^*V_{ts}[(\frac{C_3}{3}+C_4+\frac{C_9}{3}+C_{10})F^{LL}_{eP}\nonumber\\
 &+&(C_3+C_9)M^{LL}_{eP}+(C_5+C_7)M^{LR}_{eP}+(\frac{C_3}{3}+C_4-\frac{C_9}{6}-\frac{C_{10}}{2})F^{LL}_{aP}\nonumber\\
 &+&(\frac{C_5}{3}+C_6-\frac{C_7}{6}-\frac{C_8}{2})F^{SP}_{aP}+(C_3-\frac{C_9}{2})M^{LL}_{aP}+(C_5-\frac{C_7}{2})M^{LR}_{aP}]\big\} \;,\label{amp12}
\end{eqnarray}
 \begin{eqnarray}
 {\cal A}(B_s^0 \to \pi^+(K^{*-}\to)K\pi) &=& \frac{G_F} {\sqrt{2}}\big\{V_{ub}^*V_{ud}[(\frac{C_1}{3}+C_2)F^{LL}_{eK^*}+C_1M^{LL}_{eK^*}]-V_{tb}^*V_{td}[(\frac{C_3}{3}+C_4+\frac{C_9}{3}+C_{10})F^{LL}_{eK^*}\nonumber\\
 &+&(\frac{C_5}{3}+C_6+\frac{C_7}{3}+C_8)F^{SP}_{eK^*}+(C_3+C_9)M^{LL}_{eK^*}+(C_5+C_7)M^{LR}_{eK^*}\nonumber\\
 &+&(\frac{C_3}{3}+C_4-\frac{C_9}{6}-\frac{C_{10}}{2})F^{LL}_{aK^*}+(\frac{C_5}{3}+C_6-\frac{C_7}{6}-\frac{C_8}{2})F^{SP}_{aK^*}\nonumber\\
&+&(C_3-\frac{C_9}{2})M^{LL}_{aK^*}+(C_5-\frac{C_7}{2})M^{LR}_{aK^*}]\big\}  \;,\label{amp13}
\end{eqnarray}
\begin{eqnarray}
{\cal A}(B^+ \to \pi^0(K^{*+}\to)K \pi) &=& \frac{G_F} {2}
\big\{V_{ub}^*V_{us}[(C_1+\frac{C_2}{3})F^{LL}_{eK^*}+(\frac{C_1}{3}+C_2)(F^{LL}_{eP}+F^{LL}_{aP})+C_1(M^{LL}_{eP}+M^{LL}_{aP})\nonumber\\
&+&C_2M^{LL}_{eK^*}]-V_{tb}^*V_{ts}[(\frac{3C_9}{2}+\frac{C_{10}}{2}-\frac{3C_7}{2}-\frac{C_8}{2})F^{LL}_{eK^*}+\frac{3C_{10}}{2}M^{LL}_{eK^*}+\frac{3C_8}{2}M^{SP}_{eK^*}\nonumber\\
&+&(\frac{C_3}{3}+C_4+\frac{C_9}{3}+C_{10})(F^{LL}_{eP}+F^{LL}_{aP})+(\frac{C_5}{3}+C_6+\frac{C_7}{3}+C_8)F^{SP}_{aP}\nonumber\\
&+&(C_3+C_9)(M^{LL}_{eP}+M^{LL}_{aP})+(C_5+C_7)(M^{LR}_{eP}+M^{LR}_{aP})]\big\} \;,\label{amp14}
\end{eqnarray}
\begin{eqnarray}
 {\cal A}(B^0 \to \pi^0(K^{*0}\to)K \pi) &=& \frac{G_F} {2}
\big\{V_{ub}^*V_{us}[(C_1+\frac{C_2}{3})F^{LL}_{eK^*}+C_2M^{LL}_{eK^*}]-V_{tb}^*V_{ts}[(\frac{3C_9}{2}+\frac{C_{10}}{2}-\frac{3C_7}{2}-\frac{C_8}{2})F^{LL}_{eK^*}\nonumber\\
&+&\frac{3C_{10}}{2}M^{LL}_{eK^*}+\frac{3C_8}{2}M^{SP}_{eK^*}-(\frac{C_3}{3}+C_4-\frac{C_9}{6}-\frac{C_{10}}{2})(F^{LL}_{eP}+F^{LL}_{aP})-(\frac{C_5}{3}+C_6\nonumber\\
&-&\frac{C_7}{6}-\frac{C_8}{2})F^{SP}_{aP}-(C_3-\frac{C_9}{2})(M^{LL}_{eP}+M^{LL}_{aP})-(C_5-\frac{C_7}{2})(M^{LR}_{eP}+M^{LR}_{aP})]\big\} \;,\nonumber\\
\label{amp15}
\end{eqnarray}
 \begin{eqnarray}
{\cal A}(B_s^0 \to \pi^0(\bar{K}^{*0}\to)K \pi) &=& \frac{G_F} {2}
\big\{V_{ub}^*V_{ud}[(C_1+\frac{C_2}{3})F^{LL}_{eK^*}+C_2M^{LL}_{eK^*}]\nonumber\\
&-&V_{tb}^*V_{td}[(-\frac{C_3}{3}-C_4+\frac{5C_9}{3}+C_{10}-\frac{3C_7}{2}-\frac{C_8}{2})F^{LL}_{eK^*}\nonumber\\
&-&(\frac{C_5}{3}+C_6-\frac{C_7}{6}-\frac{C_8}{2})F^{SP}_{eK^*}+(-C_3+\frac{C_9}{2}+\frac{3C_{10}}{2})M^{LL}_{eK^*}-(C_5-\frac{C_7}{2})M^{LR}_{eK^*}\nonumber\\
&+&\frac{3C_8}{2}M^{SP}_{eK^*}-(\frac{C_3}{3}+C_4-\frac{C_9}{6}-\frac{C_{10}}{2})F^{LL}_{aK^*}+(\frac{C_5}{3}+C_6-\frac{C_7}{6}-\frac{C_8}{2})F^{SP}_{aK^*}\nonumber\\
&-&(C_3-\frac{C_9}{2})M^{LL}_{aK^*}-(C_5-\frac{C_7}{2})M^{LR}_{aK^*}]\big\} \;,\label{amp16}
\end{eqnarray}
where $G_F$ is the Fermi coupling constant. The $V_{ij}$'s are the Cabibbo-Kobayashi-Maskawa matrix elements.

The functions $ ( F^{LL}_{e K^*}, F^{LL}_{a K^*}, M^{LL}_{e K^*}, M^{LL}_{a K^*}, \cdots ) $
appeared in above equations are the individual decay amplitudes corresponding to different currents.
The superscript $LL$, $LR$, and $SP$  refers to the contributions from $(V-A)\otimes(V-A)$, $(V-A)\otimes(V+A)$ and $(S-P)\otimes(S+P)$ operators, respectively.
$F(M)$ describes the contributions from the factorizable (nonfactorizable) diagrams in Fig.~1 in the Ref.~\cite{prd95-056008}.
The functions $F_{e K^*}, M_{e K^*} (F_{e P}, M_{e P})$ denote the amplitudes for the $B/B_s$ meson transition into kaon-pion pair (the bachelor particle) and the functions $F_{a K^*}, M_{a K^*} (F_{a P}, F_{a P})$ represent the corresponding annihilation contributions.
Since the $P$-wave kaon-pion distribution amplitude in Eq.~(\ref{eq:phifunc}) has the same Lorentz structure as that of two-pion ones in Ref.~\cite{plb763-29},
the explicit expressions of the individual decay amplitudes can be obtained straightforwardly just by replacing the twist-2 or twist-3 DAs of the $\pi\pi$ system with the corresponding twists of the $K\pi$ ones in Eqs.~({\ref{eq:phi1}})-(\ref{eq:phi3}).



\begin{thebibliography}{199}
\bibitem{prd89-094013}
I.~Bediaga, T.~Frederico, O.~Louren\c{c}o, \prd  {\bf 89}, 094013 (2014)
\bibitem{1512-09284}
I.~Bediaga, P.C.~Magalh\~{a}es, arXiv:1512.09284 [hep-ph]
\bibitem{89-053015}
X.W.~Kang, B.~Kubis, C.~Hanhart, U.G.~Mei\ss ner,  \prd {\bf 89}, 053015 (2014)
\bibitem{prd46-1148}
T.M.~Yan, H.Y.~Cheng, C.Y.~Cheung, G.L.~Lin, Y.C.~Lin, H.L.~Yu, \prd  {\bf 46}, 1148 (1992)
\bibitem{prd55-5851}
T.M.~Yan, H.Y.~Cheng, C.Y.~Cheung, G.L.~Lin, Y.C.~Lin, H.L.~Yu, \prd {\bf 55}, 5851(E) (1997)
\bibitem{prd45-2188}
M.B.~Wise, \prd {\bf 45}, R2188 (1992)
\bibitem{plb280-287}
G.~Burdman, J.F.~Donoghue, \plb  {\bf 280}, 287 (1992)
\bibitem{prd94-094015}
H.Y.~Cheng, C.K.~Chua, Z.Q.~Zhang, \prd  {\bf 94}, 094015 (2016)
\bibitem{dalitz-plot1}
R.H.~Dalitz, Phil. Mag. {\bf 44}, 1068 (1953)
\bibitem{dalitz-plot2}
R.H.~Dalitz, Phys. Rev. {\bf 94}, 1046 (1954)
\bibitem{123-333}
R.M.~Sternheimer, S.J.~Lindenbaum, Phys. Rev. {\bf 123}, 333 (1961)
\bibitem{prd11-3165}
D.~Herndon, P.~Soding, R.J.~Cashmore, \prd  {\bf 11}, 3165 (1975)
\bibitem{BW-model}
G.~Breit, E.~Wigner, Phys. Rev.  {\bf 49}, 519 (1936)
\bibitem{prd79-094005}
B.El-Bennich, A.~Furman, R.~Kami\'nski, L.~Le\'sniak, B.~Loiseau, B.~Moussallam, \prd  {\bf 79}, 094005 (2009)
\bibitem{npb675-333}
M.~Beneke, M.~Neubert, \npb {\bf 675}, 333 (2003)
\bibitem{beneke}
M.~Beneke, talk given at The Three-Body Charmless $B$ Decays Workshop, Paris, France, 1-3 February 2006.
\bibitem{prd72-094031}
M.~Gronau, J.L.~Rosner, \prd  {\bf 72}, 094031 (2005)
\bibitem{plb727-136}
M.~Gronau, \plb  {\bf 727}, 136 (2013)
\bibitem{prd72-075013}
G.~Engelhard, Y.~Nir, G.~Raz, \prd  {\bf 72}, 075013 (2005)
\bibitem{prd84-056002}
M.~Imbeault, D.~London, \prd  {\bf 84}, 056002 (2011)
\bibitem{plb726-337}
B.~Bhattacharya, M.~Gronau, J.L.~Rosner, \plb {\bf 726}, 337 (2013)
\bibitem{plb728-579}
D.~Xu, G.N.~Li, X.G.~He, \plb  {\bf 728}, 579 (2014)
\bibitem{IJMPA29-1450011}
D.~Xu, G.N.~Li, X.G.~He, \ijmpa   {\bf 29}, 1450011 (2014)
\bibitem{prd91-014029}
X.G.~He, G.N.~Li, D.~Xu, \prd  {\bf 91}, 014029 (2015)
\bibitem{npb899-247}
S.~Kr\"{a}nkl, T.~Mannel, J.~Virto, \npb  {\bf 899}, 247 (2015)
\bibitem{plb622-207}
A.~Furman, R.~Kami\'nski, L.~Le\'sniak, B.~Loiseau, \plb  {\bf 622}, 207 (2005)
\bibitem{prd74-114009}
B.El-Bennich, A.~Furman, R.~Kami\'nski, L.~Le\'sniak, B.~Loiseau, \prd {\bf 74}, 114009 (2006)
\bibitem{APPB42-2013}
J.P.~Dedonder, A.~Furman, R.~Kami\'nski, L.~Le\'sniak, B.~Loiseau,  Acta. Phys. Polon. B {\bf 42}, 2013 (2011)
\bibitem{prd76-094006}
H.Y.~Cheng, C.K.~Chua, A.~Soni, \prd  {\bf 76}, 094006 (2007)
\bibitem{prd88-114014}
H.Y.~Cheng, C.K.~Chua, \prd  {\bf 88}, 114014 (2013)
\bibitem{prd89-094007}
Y.~Li,  \prd {\bf 89}, 094007 (2014)
\bibitem{prd87-076007}
Z.H.~Zhang, X.H.~Guo, Y.D.~Yang,  \prd {\bf 87}, 076007 (2013)
\bibitem{prd81-094033}
O.~Leitner, J.-P. Dedonder, B.~Loiseau, R. Kami$\acute{n}$ski, \prd {\bf 81}, 094033 (2010)
\bibitem{plb561-258}
C.H.~Chen, H.N.~Li,  \plb   {\bf 561}, 258 (2003)
\bibitem{prd70-054006}
C.H.~Chen, H.N.~Li, \prd  {\bf 70}, 054006 (2004)
\bibitem{prd89-074031}
W.F.~Wang, H.C.~Hu, H.N.~Li, C.D.~L\"u, \prd {\bf 89}, 074031 (2014)
\bibitem{prd97-034033}
C.~Wang, J.B.~Liu, H.N.~Li, C.D.~L\"u,  \prd {\bf 97}, 034033 (2018)
\bibitem{1803-02656}
N.~Wang, Q.~Chang, Y.L.~Yang, J.F.~Sun,  arXiv:1803.02656 [hep-ph]
\bibitem{MP}
D.~M\"uller, D.~Robaschik, B.~Geyer, F.-M.~Dittes, J.~Ho\v rej\v si,  Fortschr. Physik. {\bf 42}, 101 (1994)
\bibitem{MT01}
M.~Diehl, T.~Gousset, B.~Pire, O.~Teryaev, \prl {\bf 81}, 1782 (1998)
\bibitem{MT02}
M.~Diehl, T.~Gousset, B.~Pire, \prd  {\bf 62}, 073014 (2000)
\bibitem{MT03}
Ph.~H\"agler, B.~Pire, L.~Szymanowski, O.V.~Teryaev,  \epjc {\bf 26}, 261 (2002)
\bibitem{MN}
M.V.~Polyakov,  \npb  {\bf 555},  231 (1999)
\bibitem{Grozin01}
A.G.~Grozin,  Sov. J. Nucl. Phys.  {\bf 38},  289-292 (1983)
\bibitem{Grozin02}
A.G.~Grozin,  Theor. Math. Phys.   {\bf 69}, 1109-1121 (1986)

\bibitem{prd91-094024}
W.F.~Wang, H.N.~Li, W.~Wang, C.D.~L\"u,  \prd {\bf 91}, 094024 (2015)
\bibitem{epjc76-675}
Y.~Li, A.J.~Ma, W.F.~Wang, Z.J.~Xiao,  \epjc  {\bf 76}, 675 (2016)
\bibitem{cpc41-083105}
A.J.~Ma, Y.~Li, W.F.~Wang, Z.J.~Xiao,  \cpc {\bf 41}, 083105 (2017)
\bibitem{epjc77-199}
R.~Zhou, Y.~Li, W.F.~Wang,  \epjc {\bf 77}, 199 (2017)
\bibitem{prd97-033006}
R.~Zhou, W.F.~Wang, \prd {\bf 97}, 033006 (2018)
\bibitem{plb763-29}
W.F.~Wang, H.N.~Li,  \plb {\bf 763}, 29 (2016)
\bibitem{prd95-056008}
Y.~Li, A.J.~Ma, W.F.~Wang, Z.J.~Xiao, \prd  {\bf 95}, 056008 (2017)
\bibitem{prd96-036014}
Y.~Li, A.J.~Ma, W.F.~Wang, Z.J.~Xiao, \prd {\bf 96}, 036014 (2017)
\bibitem{npb923-54}
A.J.~Ma, Y.~Li, W.F.~Wang, Z.J.~Xiao, \npb {\bf 923}, 54  (2017)
\bibitem{1708-01889}
A.J.~Ma, Y.~Li, W.F.~Wang, Z.J.~Xiao, \prd {\bf 96}, 093011 (2017)
\bibitem{1710-00327}
A.J.~Ma, Y.~Li, Z.J.~Xiao, \npb {\bf 926}, 584-601 (2018)
\bibitem{npb924-745}
Y.~Li, A.J.~Ma, R.~Zhou, Z.J.~Xiao,  \npb {\bf 924}, 745-758 (2017)
\bibitem{ly2018}
Y.~Li, A.J.~Ma, R.~Zhou, W.F.~Wang, Z.J.~Xiao, \prd {\bf 98}, 056019 (2018)

\bibitem{prd70-092001}
B.~Aubert et al., [BABAR Collaboration], \prd   {\bf 70}, 092001 (2004)
\bibitem{prd72-072003}
B.~Aubert et al., [BABAR Collaboration], \prd  {\bf 72}, 072003 (2005)
\bibitem{prd80-112001}
B.~Aubert et al., [BABAR Collaboration],  \prd  {\bf 80}, 112001 (2009)
\bibitem{prd79-072006}
B.~Aubert  et al., [BABAR Collaboration], \prd   {\bf 79}, 072006 (2009)
\bibitem{prd78-052005}
B.~Aubert  et al., [BABAR Collaboration], \prd  {\bf 78},  052005 (2008)
\bibitem{prd83-112010}
J.P.~Lees et al., [BABAR Collaboration], \prd {\bf 83}, 112010 (2011)
\bibitem{prd75-012006}
A.~Garmash et al., [Belle Collaboration], \prd {\bf 75}, 012006 (2007)
\bibitem{prl96-251803}
A.~Garmash et al., [Belle Collaboration], \prl {\bf 96},  251803 (2006)
\bibitem{prd71-092003}
A.~Garmash  et al., [Belle Collaboration],  \prd  {\bf 71},  092003 (2005)
\bibitem{prd79-072004}
J.~Dalseno et al., [Belle Collaboration],  \prd  {\bf 79}, 072004 (2009)
\bibitem{epjc73-2373}
R.~Aaij et al., [LHCb Collaboration], \epjc {\bf 73}, 2373 (2013)
\bibitem{prl111-101801}
R.~Aaij et al., [LHCb Collaboration], \prl  {\bf 111},  101801 (2013)
\bibitem{jhep10-143}
R.~Aaij et al., [LHCb Collaboration], \jhep {\bf 10}, 143 (2013)
\bibitem{prd90-112004}
R.~Aaij et al., [LHCb Collaboration], \prd  {\bf 90},  112004 (2014)
\bibitem{prl112-011801}
R.~Aaij et al., [LHCb Collaboration],  \prl {\bf 112}, 011801 (2014)
\bibitem{prd95-012006}
R.~Aaij et al., [LHCb Collaboration],  \prd  {\bf 95},  012006 (2017)
\bibitem{jhep07-021}
R.~Aaij et al., [LHCb Collaboration], \jhep {\bf 07}, 021 (2017)

\bibitem{prd82-011502}
P.~del Amo Sanchez et al., [BABAR Collaboration],  \prd {\bf 82}, 011502(R) (2010)
\bibitem{prl97-201802}
B.~Aubert et al., [BABAR Collaboration], \prl {\bf 97}, 201802 (2006)
\bibitem{prd78-012004}
B.~Aubert  et al., [BABAR Collaboration],  \prd {\bf 78}, 012004 (2008)
\bibitem{prd84-092007}
J.P.~Lees et al., [BABAR Collaboration], \prd {\bf 84}, 092007 (2011)
\bibitem{prd75-092002}
J.~Sch\"umann et al., [Belle Collaboration],  \prd {\bf 75}, 092002 (2007)
\bibitem{prd75-092005}
C.H.~Wang et al., [Belle Collaboration], \prd {\bf 75}, 092005 (2007)
\bibitem{prd90-072009}
S.~Sato et al., [Belle Collaboration],  \prd {\bf 90}, 072009 (2014)
\bibitem{plb599-148}
P.~Chang et al., [Belle Collaboration],  \plb {\bf 599}, 148 (2004)
\bibitem{prl85-520}
S.J.~Richichi et al., [CLEO Collaboration], \prl {\bf 85}, 520 (2000)
\bibitem{prl85-2881}
C.P.~Jessop et al., [CLEO Collaboration],  \prl {\bf 85}, 2881 (2000)
\bibitem{prl89-251801}
E.~Eckhart et al., [CLEO Collaboration],  \prl {\bf 89}, 251801 (2002)
\bibitem{NJP16-123001}
R.~Aaij et al., [LHCb Collaboration],  New J. Phys. {\bf 16}, 123001 (2014)
\bibitem{jhep01-012}
R.~Aaij et al., [LHCb Collaboration],  \jhep {\bf 01}, 012 (2016)
\bibitem{prl83-1914}
M.~Beneke, G.~Buchalla, M.~Neubert, C.T.~Sachrajda,  \prl {\bf 83}, 1914 (1999)
\bibitem{npb591-313}
M.~Beneke, G.~Buchalla, M.~Neubert, C.T.~Sachrajda,  \npb {\bf 591}, 313 (2000)
\bibitem{prl96-141801}
M.~Beneke, J.~Rohrer, D.S.~Yang,  \prl {\bf 96}, 141801 (2006)
\bibitem{npb832-109}
M.~Beneke, T.~Huber, X.Q.~Li, \npb {\bf 832}, 109 (2010)
\bibitem{prd63-074009}
C.D.~L\"u, K.~Ukai, and M.Z.~Yang, \prd {\bf 63}, 074009 (2001)
\bibitem{plb504-6}
Y.Y.~Keum, H.N.~Li, A.I.~Sanda, \plb {\bf 504}, 6-14 (2001)
\bibitem{ppnp51-85}
H.N.~Li,  \ppnp {\bf 51}, 85 (2003) and references therein.
\bibitem{prd70-054015}
C.W.~Bauer, D.~Pirjol, I.Z.~Rothstein, I.W.~Stewart, \prd {\bf 70}, 054015 (2004)
\bibitem{prd74-034010}
C.W.~Bauer, I.Z.~Rothstein, I.W.~Stewart, \prd {\bf 74}, 034010 (2006)
\bibitem{npb692-232}
M.~Beneke, Y.~Kiyo, D.S.~Yang, \npb {\bf 692}, 232 (2004)
\bibitem{prd72-098501}
M.~Beneke, G.~Buchalla, M.~Neubert, C.T.~Sachrajda, \prd {\bf 72}, 098501 (2005)
\bibitem{prd72-098502}
C.W.~Bauer, D.~Pirjol, I.Z.~Rothstein, I.W.~Stewart, \prd {\bf 72}, 098502 (2005)
\bibitem{prd65-014007}
T.~Kurimoto, H.N.~Li, A.I.~Sanda, \prd {\bf 65}, 014007 (2001)
\bibitem{prd85-074004}
H.N.~Li, Y.L.~Shen, Y.M.~Wang, \prd {\bf 85}, 074004 (2012)
\bibitem{prd67-034001}
M.~Nagashima, H.N.~Li, \prd {\bf 67}, 034001 (2003)
\bibitem{1809-04754}
R.~Zhou, Y.~Li, H.N.~Li, \prd {\bf 98}, 113003 (2018)
\bibitem{plb730-336}
U.G.~Mei{\ss}ner, W.~Wang,  \plb  {\bf 730}, 336 (2014)
\bibitem{prd76-074018}
A.~Ali, G.~Kramer, Y.~Li, C.D.~L\"u, Y.L.~Shen, W.~Wang, Y.M.~Wang,  \prd {\bf 76},  074018 (2007)
\bibitem{pdg2018}
M.~Tanabashi et al., [Particle Data Group],  \prd  {\bf 98}, 030001 (2018)
\bibitem{HF2016}
Y.~Amhis et al., [Heavy Flavor Averaging Group], \epjc {\bf 77}, 895 (2017)
\bibitem{prd63-054008}
Y.Y.~Keum, H.N.~Li, A.I.~Sanda, \prd {\bf 63}, 054008 (2001)
\bibitem{epjc23-275}
C.D.~L\"u, M.Z.~Yang, \epjc {\bf 23}, 275-287 (2002)


\bibitem{npb931-79}
D.C.~Yan, P.~Yang, X.~Liu, Z.J.~Xiao, \npb {\bf 931}, 79-104 (2018)
\bibitem{prd74-094020}
H.N.~Li, S.~Mishima, \prd {\bf 74}, 094020 (2006)
\bibitem{epjc59-49}
Z.Q.~Zhang, Z.J.~Xiao, \epjc {\bf 59}, 49-66, (2009)
\bibitem{prd80-114008}
H.Y.~Cheng, C.K.~Chua,  \prd {\bf 80}, 114008 (2009)
\bibitem{prd80-114026}
H.Y.~Cheng, C.K.~Chua,  \prd {\bf 80}, 114026 (2009)
\end{thebibliography}
\end{document}